# A comparison between the styles Romanian and American students use to approach an E&M physics problem

*Cristian Raduta*[1,2]
*Gordon Aubrecht*[2]

(1) Institute of Physics and Nuclear Engineering,
Magurele, P.O.Box MG 6, Bucharest, Romania

(2) The Ohio State University, Department of Physics,
174 W. 18th Ave., Columbus, OH 43210-1106

## Abstract:
Although Physics is the same worldwide, students belonging to different learning systems as well as different cultural environments may develop different styles of approaching and reasoning out Physics problems. In a first experiment we compare student physics problem-solving styles between two different student populations: a group of typical American students (from an OSU calculus-based introductory course) and a group of Romanian students (from a second-year class at Bucharest University). We discuss one of the problems given in a short E&M survey, in which students from both populations were presented with a point charge in a region containing a uniform magnetic field. We asked students to determine the force on the charge for different initial conditions. In a second experiment, three American physics students and three Romanian physics students were given the same problem and were asked to explain loudly their answer to each question. Their answers were tape-recorded. Students' answers depend on an understanding of the Lorentz force and their general knowledge from classical mechanics. Observed similarities and differences in approach between these student groups are discussed, and our study's results described.

## I. INTRODUCTION:
Many papers have been written in the field of Physics Education (PE) about students' misconceptions related to E&M (Ref. 1-7), but few of them addressed a comparative study of students' answering styles and Physics level of students belonging to different cultural environments and learning systems. Many misconceptions about Physics (as well as about any other field) are context-dependent — depend on the learning system as well as the cultural environment in which Physics is taught. Even though Physics is the same worldwide, we all know that the way students assimilate Physics in different countries could vary quite considerable.

For example, some learning systems could put a lot of emphasis on multiple-choice type questions (MC-questions/MC-learning system), as American learning system does, while others, as some European countries do, on open-ended type questions (OE-questions/OE-learning system). Some countries introduce high level Physics classes at very early stages during student's studies — as is the case in most of the Eastern European countries, China, etc. — while others, as Americans do, postpone the introduction of high difficulty Physics classes until students reach university.

Also, some learning systems put a lot of emphasis on home-works, team-works, and high volume of work in general — as is the case with the American learning system, where Physics major students, during their undergraduate studies are kept busy with weekly home-works and projects — while other learning systems allow students to be more relaxed during their studies, the volume of home-works, projects, team-works, assigned etc. being much smaller (as is the case of Romanian learning system).

As the above examples illustrate a lot of things could be different in the context within which students, belonging to different environments and learning systems, assimilate physics. Because of these different conditions under which students study Physics in different countries we decided to do this comparison research to study in more detail the similarities and differences between two student populations belonging to two different countries. Because the two authors of this paper belong to two different countries (Aubrecht to US, and Raduta to Romania), we chose for comparison two student populations from US and Romania.

**II. RESULTS**

In a first experiment a short E&M survey was given to two student populations, one from a large midwestern university recently involved in winning a national athletic title (The Ohio State University, Columbus) and the other one from the largest university in Romania (Bucharest University, Bucharest). The American student population consisted of 74 American students most of them being sophomores majoring in electrical- and mechanical-engineering, and in computer science. The other group was formed of 52 Romanian students in the second year, all majoring in Physics.

The survey consisted of two problems, one of them being presented in Table 2.1. In this problem we wanted to test the ideas of how students in the two countries view the way charges interact with magnetic fields. Even though the number of students that have taken the survey is not statistically significant (only 52 Romanian students and 74 American students) we can still observe some similarities and differences between these two student populations.

TABLE 2.1
**Problem (E&M survey).** You have a charged particle inside a region containing a constant uniform magnetic field.
a) What is the magnetic force (*magnitude & direction*) acting on the charged particle if the initial velocity is zero? What is the trajectory of this particle?
b) What is the magnetic force acting on the charged particle if the initial speed of the charge is **v** (known, but unspecified here) and the direction is parallel to **B**? What is the trajectory of this particle?
c) What is the magnetic force acting on the charged particle if the initial speed of the charge is **v** (known, but unspecified here) and the direction is perpendicular to **B**? What is the trajectory of this particle?
d) What is the magnetic force acting on the charged particle if the initial speed of the charge is **v** (known, but unspecified here) and the angle between **v** and **B** is $\alpha$? What is the trajectory of this particle?

TABLE 2.2. American students answering each *force* part correctly and incorrectly (or had no answer)

| Task | Correct | (%) | Incorrect | % |
|---|---|---|---|---|
| a) (F = 0) | 60 | **81%** | 14 | **19%** |
| b) (F = 0) | 58 | **78%** | 16 | **22%** |
| c) (F = qvB) | 42 | **57%** | 32 | **43%** |
| d) (F = qvBsin$\alpha$) | 44 | **60%** | 30 | **40%** |

TABLE 2.3. American students answering each *trajectory* part correctly and incorrectly (or had no answer)

| Task | Correct | (%) | Incorrect | (%) |
|---|---|---|---|---|
| a) (at rest) | 32 | **43%** | 42 | **57%** |
| b) (straight line) | 24 | **32%** | 50 | **68%** |

| c) (circle) | 6 | **7%** | 68 | **93%** |
| d) (spiral) | 0 | **0%** | 74 | **100%** |

TABLE 2.4. Romanian students answering each *force* part correctly and incorrectly (or had no answer)

| Task | Correct | (%) | Incorrect | % |
|---|---|---|---|---|
| a) (F = 0) | 23 | **44.2%** | 29 | **55.8%** |
| b) (F = 0) | 19 | **36.5%** | 33 | **63.5%** |
| c) (F = qvB) | 26 | **50%** | 26 | **62.5%** |
| d) (F = qvBsinα) | 16 | **25%** | 36 | **75%** |

TABLE 2.5. Romanian students answering each *trajectory* part correctly and incorrectly (or had no answer)

| Task | Correct | (%) | Incorrect | (%) |
|---|---|---|---|---|
| a) (at rest) | 19 | **36.5%** | 33 | **63.5%** |
| b) (straight line) | 24 | **46.1%** | 28 | **53.9%** |
| c) (circle) | 17 | **32.7%** | 35 | **67.3%** |
| d) (spiral) | 10 | **19.2%** | 42 | **80.8%** |

TABLE 2.6 Results for both groups of students for the force part from question c).

| | a lot of words / rich in explanations | | | Some words / some explanations | | | Very few words (if some) / almost no explanations | | |
|---|---|---|---|---|---|---|---|---|---|
| | Correct answer | Wrong answer | Total | Correct answer | Wrong answer | Total | Correct answer | Wrong answer | Total |
| **US students** | 7 | 4 | **11 (14.8%)** | 32 | 23 | **55 (74.3%)** | 3 | 5 | **8 (10.8%)** |
| **Rom students** | 18 | 10 | **28 (53.8%)** | 4 | 6 | **10 (19.2%)** | 4 | 10 | **14 (26.9%)** |

In tables 2.2, 2.3, 2.4 and 2.5 could be seen the results obtained by the two student populations. The results drawn from this problem are pretty alarming especially if we take into account that this two-question survey — with two easy standard problems — was given just two weeks before the end of the quarter, when one would expect students to be pretty comfortable with the main concepts of E&M (Note: we discussed only the magnetic field problem here. We also asked a Gauss' Law problem). In table 2.6 we present another way of examining their answers function of the number of words used in answering question c).

**Observations:**
Some students (both Americans and Romanians) seem to not know what the difference between a scalar and a vector is, misconception well documented in physics education research (PER). In many cases, students would have a scalar in one side of the equation and a vector (or a vector product) in the other side. Even if a student made this mistake we decided to include him as having a correct answer (if everything else was correct).

Even we instructed all the students to be as explicit as possible when answering their questions — to use "words" instead of just a few formulas — a lot of American students just wrote some formulas.

The trajectory question seemed to be pretty difficult for most of the students (both Americans and Romanians). Only 7% of the American students and 25% of the Romanian students realized that a charged particle coming into a magnetic field with a velocity perpendicular on the field will have a circular motion. These results are consistent with the findings of Bagno et al.'s (Ref 4).

None of the American students gave a correct answer for the trajectory (a spiral) when the charged particle is coming under an angle ≠ 90 degrees in a magnetic field; only 10 Romanian students (19.2%) identified correctly the trajectory as being a spiral.

For question a), even it was very easy to find the right trajectory, only 43% of the American students and 36.5% of the Romanian students gave the correct answer. The same happened with the trajectory question from part b) which was slightly more difficult than that from part a). Only 32% of the American students and 46.1.% of the Romanian students realized that a particle with initial velocity **v** will have a straight line trajectory if no force will act on it.

These results point again to the difficulty of the students to deal simultaneously with concepts from both Mechanics and E&M that is well documented in other studies (Ref. 4, 7, 8).

A lot of the students (mostly Americans) wrote the Lorentz force correctly up to the electric charge q.

When trying to get the direction of the force, most of them (Americans) were talking like: "*the direction can be found using the right-hand rule.*" But we haven't seen one statement talking about vector product. It seems that American students are used more to the right hand rule, while Romanian students more to the vector product. None of the Romanian students had any statement related to the right hand rule.

### III. STYLE ANALYSIS
In Table 3.1 bellow we present two representative "good answers" from two students from the two populations considered in the first experiment. As one can easily see, their answers look very different. While the Romanian student (whose answer was presented) is writing more words, trying to be as explicit as he can answering each question, the American student is very brief in his answers, writing mostly formulas (and this held true to a great extent for most Romanian and American students who gave reasonable good answers for the problems). Looking at American student's answer we notice that he's trying to be as concise and direct as possible, leaving away unnecessary words that could easily be implied -- as in the statement, "⊥ *to magnetic field*", that clearly means that the trajectory of the particle is perpendicular to the direction of the magnetic field. In fact, as one can see in Table 3.1 bellow, the American student is using basically the minimum number of formulas, words or symbols necessary for answering each question (and justifying an answer), while the Romanian student is trying to explain in words each step in the reasoning process needed for solving the problem.

On the other hand, analyzing the student category from each population, who didn't know how to answer a question, American students seemed to have more courage (confidence) in approaching the questions, throwing on the paper some words or formulas (even though not necessarily related to the questions), while most of the Romanian students from this category left the page blank.

TABLE 3.1
*Romanian student representative (good) answer:*
a) The magnetic force for a charge in an uniform field is: **f**=q**v**x**B**. If v=0, than f=0, and it will not be accelerated in the field, hence we can't speak of direction of the force, but we can say that the magnitude is always zero.

b) **v**∥**B**, v≠0, **f**=q**v**x**B**=qv**B**sinα; **v**∥**B**=0→ α=0 →sinα=0; →**f**=0; Hence the trajectory is a straight line parallel the lines of magnetic field. The equation of the motion will be:
$$x=x(0)+vt, \text{ where } v=ct.$$
c) if **v**⊥**B**, then α=90, sinα =1. The trajectory of the particle will be a circle perpendicular to the magnetic field lines. The magnitude of the force is f=qvB, and the direction is that of the radius of the circle pointing towards the center of the circle.

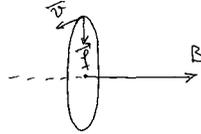

d) ‹(**v**;**B**)=α, is the superposition of the two previous cases, and the trajectory of the particle will be a helicoidal one, with parameters radius and step:
$$\text{step}=v(\parallel) T; \text{ radius}=f(v(\perp); m)$$

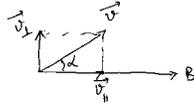

where m is the mass of the particle. The magnitude of the magnetic force is f=qvBsinα and the direction is always perpendicular to the trajectory.

*American student representative (good) answer:*
a) **F**(L)=q**v**x**B**; if v=0, then **F**(L)=0;   ⊥to magnetic field
b) **F**(L)=q**v**x**B**=qv**B**sinθ =qvx**B**sin0=0;   across magnetic field
c) **F**(L)=q**v**x**B**=qv**B**sinθ =qv**B**sin90=qv**B**,   ⊥ to v and B;
d) **F**(L)=qv**B**sinθ; ⊥ to v and B,   ⊥ to v and B

In a second experiment (that was done in Romania) three American students and three Romanian students were interviewed and tape-recorded in the same time. They were asked to explain loudly the solution to each part of the problem from Table 2.1, trying to observe style differences between these two small student populations. The results for part c) are presented in Table 4.7 bellow.

TABLE 4.7 Different characteristics of students' answers given to the problem above.

| Student | # of words used (only to part c)) | Uses short phrases vs. long phrases | Tone, confidence | # of times student talk about something else |
|---|---|---|---|---|
| Mike (Am. student) | 36 | short phrases | Const. tone, confident | 0 |
| Barney (Am. student) | 40 | short phrases | Const. tone, less confident | 2 |
| Timothy (Am. student) | 68 | long phrases | Const. tone, little gesture, confident | 1 |
|  |  |  |  |  |
| Artenie (Rom. student) | 89 | long phrases | Lot of gesture, sinusoidal tone, confident | 4 |
| Ilie (Rom. student) | 58 | long phrases | Const. tone, less confident | 1 |
| Cristi (Rom. student) | 67 | long phrases | Some gesture, sinusoidal tone, some confidence | 3 |

The results from the videotaped interviews do not contradict the results obtained in the written survey. As in the written surveys, American students seem to be overall more confident on their answers than Romanian students did, answering each question using smaller number of words than their omologs from Romania. Also, American students seemed to be more focused ("pure activity mode" vs. "mixed activity mode") on the interview than Romanian students did, talking fewer times during the interview about something else not related with the interview.

**IV) POSSIBLE EXPLANATIONS FOR THE STYLE DIFFERENCES**
In trying to understand the differences in styles of these two student populations, we have to examine the cultural and learning system differences under which these students have been exposed during their lifetime.

*Cultural differences*
i) Even though the two countries have a lot of cultural similarities, — American culture vs. European countries' cultures being much more similar than US or European culture vs. Asian countries' cultures or any other part of the world — there are still some important differences that could be reflected also in their problem solving styles. While American Anglo-Saxon culture is a mature business-oriented one in which efficiency, strict rules and highly specialized professionals are highly desired, Romanian Latin culture is quite different from this point of view, having a quite different path of history, habits and social interactions. The rhythm of life in a faced paced business-oriented enviroment in which the expression "time is money" is quite representative for the whole culture is quite different from the one from Romania or other East European countries.

*Learning system*
ii) The two countries' learning systems are also quite different:

a) While in Romania a big number of classes of Mathematics, Physics and Chemistry are taught in most High-Schools, in US few High-Schools deliver such classes, for most of American students the first advanced classes of Physics being taken in University. We believe this is an important factor that could be directly correlated with business rationales.

b) While in Romania, and most European countries the university studies are basically free (or almost free), in US college education could cost quite a lot, especially if you are an out-of-state student. This is also an important difference between the two learning systems which brings a business component into the learning process. "Learning" ("knowledge") becomes like a business package product ready for sale for the annual tuition. For example, for an out-of state student at OSU the tuition is around 20,000$/yr. while at Bucharest University you pay just some symbolic taxes (around 30$/yr.). This has two important consequences:

1) Students have less time for studying (or less leisure time), because a lot of them, in order to be able to pay the tuition and other living expenses, are forced to work part-time besides school.

2) There is a psychological subtle effect involved because of this big tuition American students have to pay (and implicitly loans to be paid back). The learning process is transformed into *an initial (business) investment* that should bring a good return (material) in the long-run. This forces the students to become more efficient at their exams ("check-ers"), being more interested in taking good grades than in really understanding and thinking at the knowledge they have been learning.

3) One feature of the American learning system seen in primary school through the university level (and, for many fields of study, even at the graduate level) is an emphasis on multiple-choice (MC) questions. Among the other global learning systems, Americans seem uniquely reliant on MC questions. Most other learning systems, including Romanian one, do use the MC questions to some extent — in many cases imitating the American system — but few use them to the same extent as they are used in the US.

Many papers have been written about the use of the MC questions in PER (Ref. 9-12) and in education research in other areas. In most of them, the researchers were working to develop effective multiple-choice tests intended to be able to evaluate and compare instructions that are delivered to large populations. The question of which sort of test (objective, essay, etc.) to use during exams has been discussed by many authors (see, for example, Refs. 13-16). There have been both, emotional and substantive appeals for the use of objective tests, and equally forceful statements opposed to objective tests. We believe that, if wisely combined and designed, both types of tests can be very useful.

Among other consequences that might have, if used extensively, MC questions help students feel more confident of their "knowledge" — a paradox! — eliminating the confusion and uncertainty in choosing from among many other potential answers (created or memorized). For a MC question, on an exam, there is clearly only one correct answer, which must be among those written down by the instructor. Once it is chosen, that's it, the mental check-off is done, the job completely finished. The student can confidently move on to the next question. In the long-run, this could contribute to the self-confidence and transparency (and sometimes obstinacy and unwillingness to listen) many find characteristic of American culture; and our findings are consistent with this.

TABLE 3.2 Cultural and learning differences between US and Romanian

| Country | Physics classes in high-school? | Tuition | Type of exam questions | Student's leisure time | Culture-type |
|---|---|---|---|---|---|
| **US** | No (in most high-schools) | Yes | More reliant on MC-type questions | Less | Business-oriented |
| **Romania** | Yes (in most high-schools) | No | More reliant on OE-type questions | More | Semi-Business-oriented (more people-oriented) |

**CONCLUSSIONS**

In this paper we analyzed some of the style differences in approaching and reasoning out Physics problems, students belonging to different cultural environments or learning systems might develop. Even though, in terms of the physics answers they gave, the two student populations considered are very similar – American doing slightly better – their styles differ quite considerable. While Romanian students wrote more words than their omologs from US, trying to explain in words each step in the reasoning process needed to answer each question, American students were very brief and concise in their answers. During the videotaped interviews, Romanian students talked more, made more gesture and seemed less confident in their answers than their American homologues. In the same time, American students seemed to be more focused on the interview than Romanian students did, talking very few times during the interview about something else than what were expected. So, they seem to be more *result-driven* than their collegues from Romania. We tried to explain all these style differences through the cultural and learning-system differences that definitely exist between the two countries considered.